\documentclass[conference]{IEEEtran}
\IEEEoverridecommandlockouts
\usepackage{cite}
\usepackage{amsmath,amssymb,amsfonts}
\usepackage{algorithm, algorithmic}%
\usepackage{graphicx}
\usepackage{multicol}
\usepackage{textcomp}
\usepackage{xcolor}
\def\BibTeX{{\rm B\kern-.05em{\sc i\kern-.025em b}\kern-.08em
    T\kern-.1667em\lower.7ex\hbox{E}\kern-.125emX}}
\begin{document}

\title{Decentralized shape formation and force-based interactive formation control in robot swarms\\
}

\author{\IEEEauthorblockN{Akshaya C S}
\IEEEauthorblockA{\textit{Instrumentation and Control Engineering} \\
\textit{NIT - Tiruchirappalli}\\
Tiruchirappalli, India \\
akshayasub@gmail.com}
\and
\IEEEauthorblockN{Karthik Soma}
\IEEEauthorblockA{\textit{Instrumentation and Control Engineering} \\
\textit{NIT - Tiruchirappalli}\\
Tiruchirappalli, India \\
soma.nitt@gmail.com}
\and
\IEEEauthorblockN{Visweswaran B}
\IEEEauthorblockA{\textit{Electrical and Electronics Engineering} \\
\textit{NIT - Tiruchirappalli}\\
Tiruchirappalli, India \\
visweswaran21@gmail.com}
\and
\IEEEauthorblockN{Aditya Ravichander}
\IEEEauthorblockA{\textit{Instrumentation and Control Engineering} \\
\textit{NIT - Tiruchirappalli}\\
Tiruchirappalli, India \\
aditya.ravichander@gmail.com}
\and
\IEEEauthorblockN{Venkata Nagarjun PM}
\IEEEauthorblockA{\textit{Electrical and Electronics Engineering} \\
\textit{NIT - Tiruchirappalli}\\
Tiruchirappalli, India \\
nagarjun.nitt@gmail.com}
}

\maketitle

\begin{abstract}
Swarm robotic systems utilize collective behaviour to achieve goals that might be too complex for a lone entity, but become attainable with localized communication and collective decision making. In this paper, a behaviour-based distributed approach to shape formation is proposed. 

Flocking into strategic formations is observed in migratory birds and fish to avoid predators and also for energy conservation. The formation is maintained throughout long periods without collapsing and is advantageous for communicating within the flock. Similar behaviour can be deployed in multi-agent systems to enhance coordination within the swarm. Existing methods for formation control are either dependant on the size and geometry of the formation or rely on maintaining the formation with a single reference in the swarm (the leader). These methods are not resilient to failure and involve a high degree of deformation upon obstacle encounter before the shape is recovered again. To improve the performance, artificial force-based interaction amongst the entities of the swarm to maintain shape integrity while encountering obstacles is elucidated.      
\end{abstract}


\section{Introduction}
Swarm robotics is a field of using multiple agents to coordinate and complete tasks without a central authority. It derives from biological swarms in nature, which use emergent behaviour to achieve complex tasks.The intent behind collective behavior is to integrate individual behavior into an aggregate pattern by utilizing the synergy of the group. There is no dearth of examples for collective behaviour in nature - flocking in birds, foraging in ants, hive building in bees, synchronisation in fireflies, etc. Several biologically inspired algorithms mimic swarm behaviour in artificial multi-agent systems to improve the robustness of the system. It is interesting to note that in many cases, the individual entity of the swarm has limited capability - be it in terms of sensing, decision making or motion and actuation, but the swarm circumvents these limitations by integrating individual resources with the help of social behaviours.  

For instance, during long migrations, birds which have limited field of vision are known to travel in formations to use the ‘collective vision’ of the group for enhanced coordination and to be wary of predators and other dangers. Upon analysing the energy distributions of the 'V' shaped formation, scientists have inferred that the shape minimises the overall energy currency of the swarm. This is because the birds position themselves in regions with higher lift\cite{b1}.

It is worth noting that while the birds might not be aware of the broader perspective of energy minimisation, by achieving the local goal of applying the least exertion, they could attain global energy minimisation. It is also noticed that when the leader of the formation at the tip of the arrow is tired, there is a social structure to elect a new leader to take its place so that it can fall behind and take advantage of the aerodynamic lift created by its predecessors\cite{b2}.  

Drawing inspiration from this to emulate collective behaviours, we propose a decentralised method for shape formation in robot swarms and also devise swarm-wide strategies to avoid obstacles while maintaining the formation.
 
\section{Related Work}
Biological swarms in nature have inspired research in computational methods such as Particle Swarm Optimisation (PSO) \cite{b3} and Ant Colony Optimisation (ACO) \cite{b4}. Popular swarm behaviours that have been widely researched include aggregation, flocking, construction, foraging and synchronisation. 

Pattern formation has been researched extensively to demonstrate the cornerstones of collective behaviour - fault tolerance, immunity to a single point of failure and scalability. In \cite{b5} the shape is input as a point cloud and each waiting bot chooses two parents as reference to form the shape progressively. The famous Lennard Jones potential has been used for bot kinematics in aggregation-based formation strategies in \cite{b6} and \cite{b7}. In \cite{b8}, the task is divided into goal selection and motion planning to achieve pattern formation in large swarms. The hardware experiment with a swarm of 1024 robots in \cite{b9} successfully converged into the required shapes. Each bot, called a Kilobot \cite{b10} is small, simple and primitive with light and distance sensors, but in a swarm their abilities are multiplied by many folds. 

While the works discussed above adopt an additive strategy for shape formation, in \cite{b11}, a subtractive approach is illustrated. Initially the kilobots are assembled in a grid. Each kilobot in the swarm uses phototaxis and multilateration to decide whether it is part of the shape or not. Since pattern formation involves the movement of bots from random locations in the environment to a relatively smaller subspace, there is a chance of collision between two moving robots. To prevent this, a field-based method is discussed in \cite{b12} and \cite{b13} for static obstacles. 

Reaching a goal point while maintaining the formation can prove to be useful in exploration and search and rescue operations \cite{b14}. An interesting parallel with swarms of fish is drawn in \cite{b15} and a mechanical solution is realised using a spring-mass-damper system to preserve the shape while encountering obstacles.

Formation control and obstacle avoidance using evolutionary algorithms is proposed in \cite{b16} while \cite{b17} performs self- healing upon encountering obstacles. In \cite{b18}, the entire swarm dynamically reaches a consensus on the azimuth to avoid obstacles while \cite{b19} employs trajectory generation and predictive control to reach the goal without colliding with obstacles. Reference \cite{b20} explores the possibility of electing a leader to survey the obstacle so as to fragment the shape, overcome it and re-join with the main shape after crossing.

The objective of \cite{b21} is similar to our problem statement, however every bot knows the final coordinates and moves independently to the destination after forming the shape. All the bots rely on global co-ordinates and bot-to-bot communication is activated only when the positional error with respect to the reference is greater than a threshold. In this paper, a more intuitive method reliant on neighbor interactions is suggested, with only one bot knowing the destination at any given time, reducing the sensory overhead for the rest of the bots.

\section{Methodology}
The objective of the paper is to achieve the following tasks: shape formation, formation control during motion, obstacle avoidance with obstacle size discrimination. 

\subsection{Notation}
The constants and variables used in the equations to follow are defined in Table \ref{tab1}. 
\begin{table}[htbp]
\caption{Definition of the variables used}
\begin{center}
\begin{tabular}{|c|c|}
\hline
\textbf{Variable}& \textbf{Definition} \\
\hline
$\overrightarrow{F}_{collision}$&{Force generated to avoid bot-bot collision}\\
\hline
$k_{collision}$&{Proportional constant for collision avoidance}\\
\hline
$r_{ij}$&{Distance between bots i and j}\\
\hline
$r_o$&{Minimum threshold distance for collision }\\
\hline
$\alpha_{ij}$&{Azimuth of bot j with respect to i}\\
\hline
$\overrightarrow{x}, \overrightarrow{y}$&{Reference axes in the local Cartesian plane}\\
\hline
$k_o$&{Proportional constant for reference force}\\
\hline
$r_{0,parent}$&{Reference distance from parent}\\
$r_{t,parent}$&{Dynamic distance from parent}\\
\hline
$\alpha_{0,parent}$&{Reference azimuth of parent}\\
$\alpha_{t,parent}$&{Dynamic azimuth of parent}\\
\hline
$\overrightarrow{F}_{goal}$&{Force on 'leader' to reach the goal}\\
\hline
$\overrightarrow{r}_{goal}$&{Vector of the goal from origin}\\
$\overrightarrow{r}_{position}$&{Position vector of the bot}\\
\hline
$\overrightarrow{F}_{a,t}$&{Dynamic attractive force w.r.t parent}\\
\hline
$\overrightarrow{F}_{net}$&{Net attractive force on the bot}\\
\hline
$r_{sensor}$&{Obstacle distance returned by the sensor}\\
\hline
$\alpha_{sensor}$&{Angle of the sensor}\\
\hline
n&{Total number of ultrasonic sensors on the bot}\\
\hline
m&{Proportionality constant for obstacle force}\\
\hline
\end{tabular}
\label{tab1}
\end{center}
\end{table}

\subsection{Shape Input and Pre-Processing}
\begin{figure}[b]
\centerline{\includegraphics[width = 80mm]{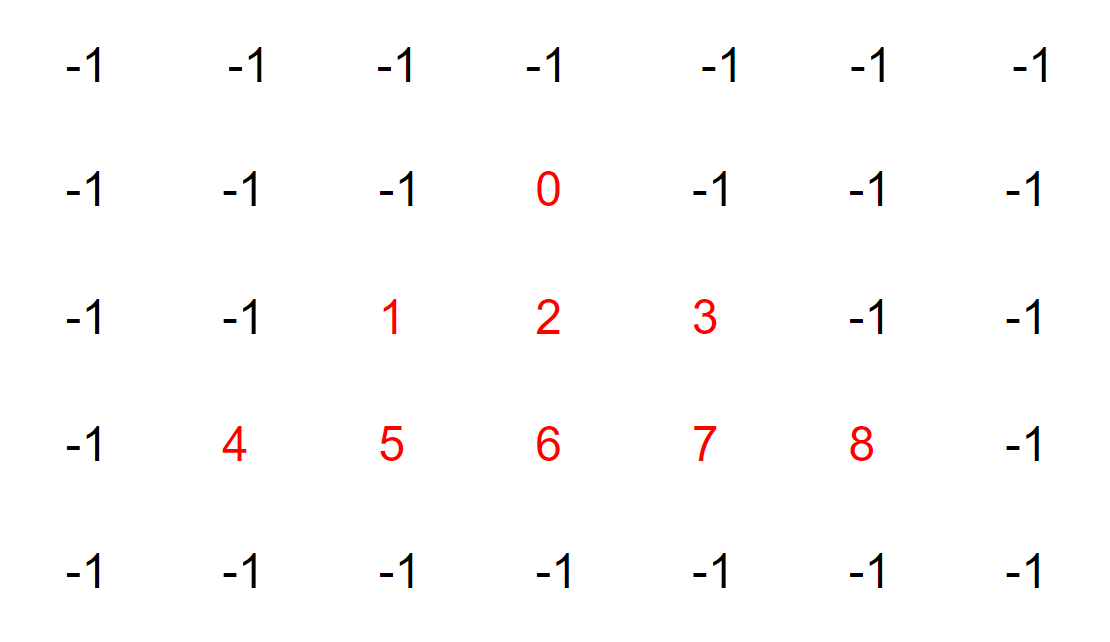}}
\caption{Shape matrix for a triangle.}
\label{fig:1}
\end{figure}

The shape is input in the form of an integer matrix. Every integer on the matrix corresponds to a label on the shape and the labels range from 0 to n, where n is the total number of bots required to form the shape. The elements which are not part of the shape are given a label of -1. The input matrix is similar to a bitmap described in \cite{b21}, but includes integer labels for assignment instead of just 0s and 1s as shown in Fig.~\ref{fig:1}. The matrix is processed into a table indexed by the labels containing information about the relative angles and distances of the shape.

\begin{table}[b]
\caption{Row for label 0 in shape table}
\begin{center}
\begin{tabular}{|c|c|c|c|}
\hline
\textbf{Label/}&\multicolumn{3}{|c|}{\textbf{Grid Information}} \\
\cline{2-4} 
\textbf{Index} & \textbf{\textit{Neighbor Labels}}& \textbf{\textit{Angle(radian)}}& \textbf{\textit{Distance(m)}} \\
\hline
0&{1, 2, 3}&{$-3\pi/4$, $\pi/2$, $-\pi/4$}&{1.4, 1.0, 1.4}\\
\hline
\end{tabular}
\label{tab2}
\end{center}
\end{table}
Each element of the shape matrix is scanned and the 8 elements surrounding it are considered as neighbours. Making use of the information on every 3x3 local grid as shown in Table \ref{tab2}, the relative angles and distances are mapped out in the shape table for later use.

\subsection{Label Allotment and Movement}
Once the table is ready, the bots exchange information about their coordinates with respect to the origin in the arena (0,0,0).
\begin{figure}[t]
\centerline{\includegraphics[width = 80mm]{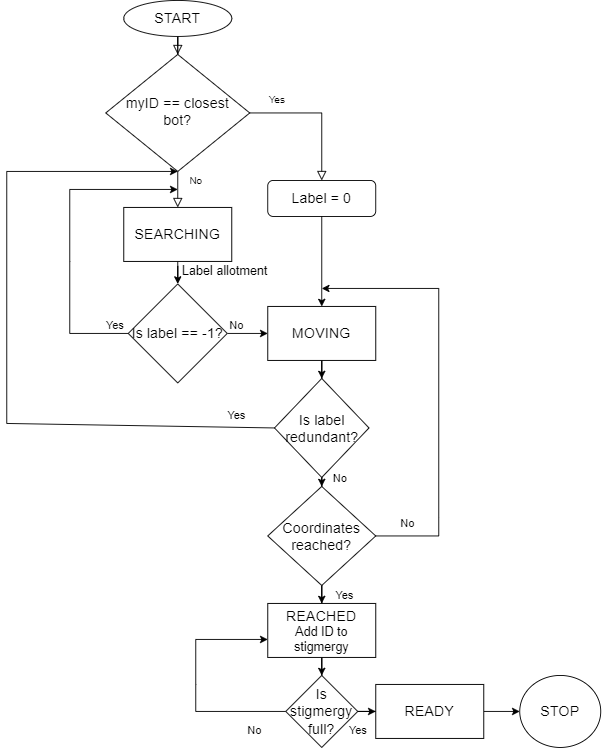}}
\caption{Flowchart describing the progression of states}
\label{fig:2}
\end{figure}
\begin{itemize}
\item Searching: All bots are in state ‘searching’ initially and have a label of -1. Based on local interactions, the bot closest to the origin is determined and is allotted the label of 0. This bot is the seed bot and it moves to state ‘moving’. Now it broadcasts the locally available labels inside its 3x3 grid to the rest of the swarm in the form of broadcasts. The rest of the swarm in state ‘searching’ actively listens for available labels and appends these labels to a list. 

It also checks if the labels are viable by listening to neighbouring bots in other states. If a label in the list is already taken by a neighbour, the value of that element in the list is changed to -1. The last non-negative label is the allotted label, but it is subject to change in the next state as described later. If there is a non-negative label in the list, it becomes the bot’s label and it moves to ‘moving’ state. If not, it remains in ‘searching’.       

\item Moving: Upon receiving a label, the bot moves towards its coordinates relative to the position of its ‘parent label’ obtained from the shape table while also actively broadcasting labels in its local grid. This way all the labels in the shape matrix are covered in a progressive grid-by-grid manner. 

The bot also checks if the same label is being shared by other bots. In case of a redundant label, the bot with a higher ID gets the label while the other bots go back to ‘searching’.

Since the  dispersed bots will now move to form a fairly dense shape, there are bound to be collisions. To avoid this, a repulsive force \eqref{eq1} is activated for neighbours closer than a threshold distance \(r_{o}\). The net force vector is the sum of repulsive forces of the individual neighbour forces. 
\begin{equation}
\overrightarrow{F}_{collision} = \frac{-k_{collision}}{(r_{ij} - r_o)^2}(cos(\alpha_{ij})\overrightarrow{x} + sin(\alpha_{ij})\overrightarrow{y})\label{eq1}
\end{equation}

In such a case, the task changes from reaching the desired position to avoiding collision as a combined force could lead to an equilibrium point.

\item Reached: Upon reaching the coordinates obtained from its parent’s index in the shape table, the bot turns and orients itself along the same direction as the leader. Upon reaching the desired orientation, it adds its ID to a virtual stigmergy structure to reach consensus on the completion of the shape. The virtual stigmergy in Buzz is a distributed key - value structure that enables swarm-wide communication\cite{b23}.

\begin{algorithm}[hbt!]
\caption{Task completion consensus}\label{alg:one}
\begin{algorithmic}
\IF{$state == reached$} 
    \IF{$|orientation - clearance|> reference$} 
     \STATE rotate
     \ENDIF
\ELSE
    \STATE put ID in stigmergy
    \IF{$stigmergy.size() == threshold$}
        \STATE $state \gets ready$
\ENDIF
\ENDIF 
\end{algorithmic}
\end{algorithm}
This stigmergy structure acts as a barrier for task completion. All the bots are required to reach consensus on shape completion. When the desired threshold for task completion is reached, the bots move on to state ‘ready’.

\item Ready: There is a timeout in this state before transitioning into the next task. The entire shape is analysed by local interactions by exchanging relative azimuth values among neighbors. From these interactions the bots on the boundaries are identified and are given roles of ‘leftmost’ and ‘rightmost’. Each bot evaluates its surroundings in a limited way, but upon integrating these observations using stigmergy, the border bots are identified. These are bots which are most affected while encountering obstacles, hence they need to be identified and known to the entire swarm. From the set of boundary bots, a leader bot is elected by evaluating the relative azimuths of all the boundary bots.

To maintain the shape, each bot with the exception of the ‘leader’ chooses a parent. The relative azimuth and distance are considered while choosing a parent. To avoid deadlock, a common parents table with ID of all the bot’s parents is broadcast via local stigmergy. Now each bot calculates an attractive force \eqref{eq:2} with respect to its parent in the structure.
\begin{equation}
\overrightarrow{F}_{ref} = k_o{r_{0,parent}^2}(cos(\alpha_{0,parent})\overrightarrow{x} + sin(\alpha_{0,parent})\overrightarrow{y})\label{eq:2}
\end{equation}
This force vector is shifted by $\pi$ radians to act as a reference counter-force when the swarm starts moving. Once the timeout is over, the consensually elected leader bot experiences an attractive force towards the goal \eqref{eq:3} .
\begin{equation}
\overrightarrow{F}_{goal}  \propto \overrightarrow{r}_{goal} - \overrightarrow{r}_{position}\label{eq:3}
\end{equation}
The rest of the bots calculate the attractive force with respect to their parents as the leader begins to move towards the destination.
\begin{equation}
\overrightarrow{F}_{a} = k_o{r_{t,parent}^2}(cos(\alpha_{t,parent})\overrightarrow{x} + sin(\alpha_{t,parent})\overrightarrow{y})\label{eq:4}
\end{equation}
The role of the leader bot is to a) avoid obstacles b) reach the goal. Meanwhile the rest of the swarm does not know the destination coordinates. Their task is to a) Maintain formation b) Avoid obstacles.
\end{itemize}

\subsection{Formation Control}
Each child bot dynamically calculates the force (equation 1) with respect to its parent. The initial reference force is subtracted from this force and the net attractive force drives the bot towards maintaining its position relative to its parent similar to the network topology of a directed graph. 
\begin{equation}
\overrightarrow{F}_{net} = \overrightarrow{F}_{a} - \overrightarrow{F}_{ref}\label{eq:5}
\end{equation}

\subsection{Obstacle Avoidance}
Each Khepera IV\cite{b22} bot comes with 5 ultrasonic sensors placed over a range of $\pi$ radians. The obstacles are thought of as repelling entities that the bot should avoid. The ultrasonic sensor readings are summed up to produce a repulsion field on the bot.
\begin{equation}
\overrightarrow{F}_{obs} = \sum_{sensor}^{n} \frac{m}{r_{sensor}}(cos(\alpha_{sensor})\overrightarrow{x} + sin(\alpha_{sensor})\overrightarrow{y})\label{eq:6} 
\end{equation}
where
\begin{equation}
m \propto |{\overrightarrow{F}_{goal}}|\label{eq:7}
\end{equation}

It is important for the bot to discriminate between an obstacle in its environment and a neighbouring bot. For this, the bot gets the relative azimuth of its neighbours from local broadcasts and does not factor in the ultrasonic sensor values corresponding to those angles into the force vector addition.

\begin{algorithm}[hbt!]
\caption{Obstacle repulsion calculation}\label{alg:two}
\begin{algorithmic}
\FOR{$a=-\pi/2$ to $+\pi/2$}   
\IF{sensor at angle a detects obstacle}
    \STATE listen to neighbours’ broadcast 
    \STATE get neighbours’ azimuth
	\FORALL{neighbours}
		\IF{$|azimuth - clearance| > a$} 
			\STATE calculate repulsive force vector\\
                \STATE add to the force accumulator 
            \ELSE 
	        \STATE repulsive force component = 0
	    \ENDIF
	\ENDFOR
    \ENDIF
\ENDFOR
\end{algorithmic}
\end{algorithm}
The swarm also discriminates between obstacle sizes and dynamically adopts a strategy which is best suited to its environment with the least disturbance to the overall shape.

In case of large obstacles which could disturb the formation, for the perimeter bots, the repulsion force becomes dominant at a certain distance r. This causes the shape to shrink in size from that corner. Once the sensors no longer detect any obstacle, the repulsion force becomes zero and the bot is only acted upon by the attractive force $\overrightarrow{F}_{net}$  \eqref{eq:5}. 

Thus, different weights are given dynamically to the forces depending upon the environment. There is a trade-off between maintaining shape integrity and avoiding the obstacle.    
   
For smaller obstacles relative to the size of the formation, the optimum strategy would be for the formation to fragment and re-join with the main shape after the obstacle is passed. 
The bot decides on the course of action from the ultrasonic sensor fusion. 

Upon encountering a smaller obstacle, wall-following behaviour is done only for the affected bot and the rest of the swarm is undisturbed. Upon reaching the destination all the forces are reduced to zero and the final state is reached with some amount of deformation in the shape. 

\section{Results}
The simulation is executed on ARGoS - a multi robot simulator \cite{b23}.   
The controller logic is written in Buzz \cite{b24} and this controller script is responsible for the behaviour of the bots. The script runs simultaneously in all the bots during execution.

\begin{figure}[htbp]
\centering
\includegraphics[width = 80mm]{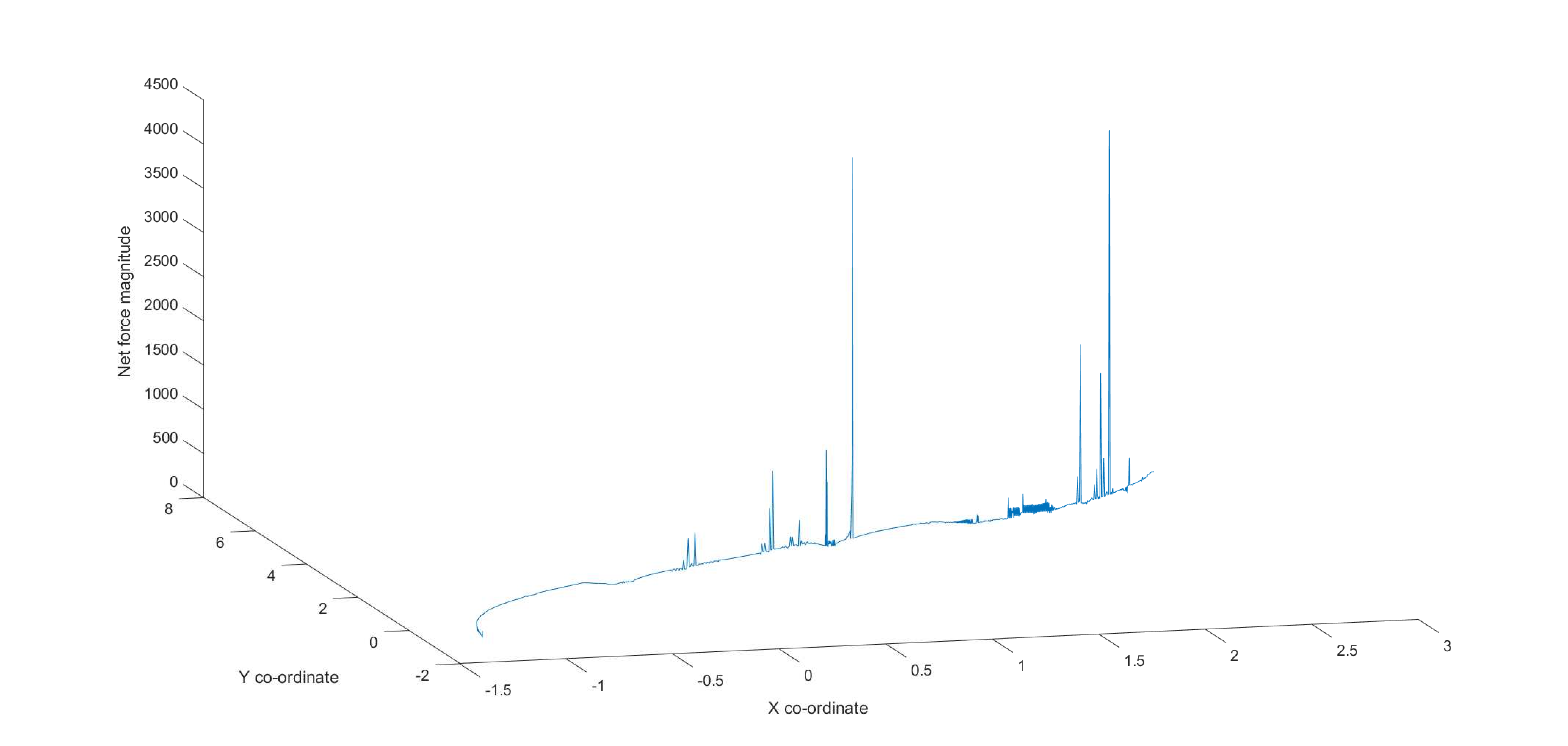}
\caption{Sum of the net force on the swarm against various positions in the trajectory}
\label{fig:3}
\end{figure}

\begin{figure}[htbp]
\centering
\includegraphics[width = 80mm]{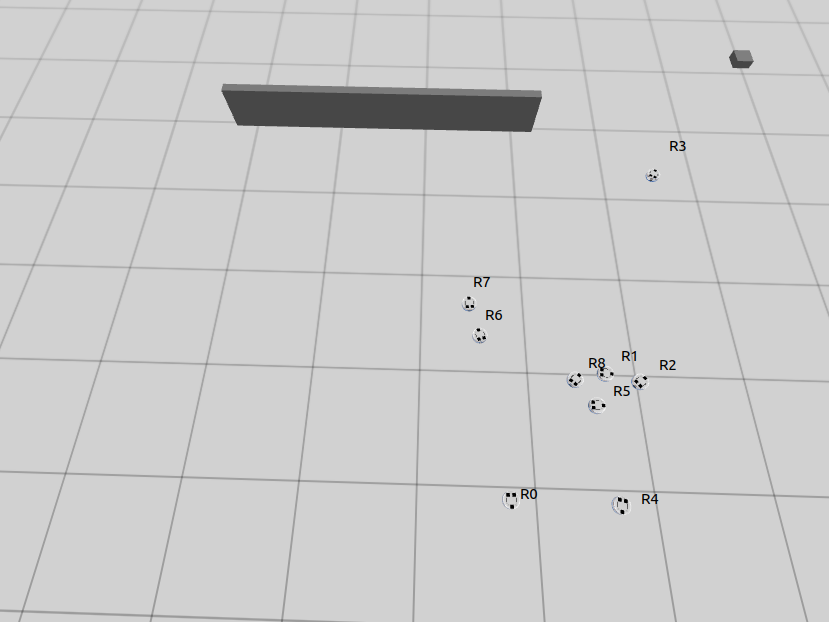}
\caption{t = 0}
\label{fig:4}
\centering
\includegraphics[width = 80mm]{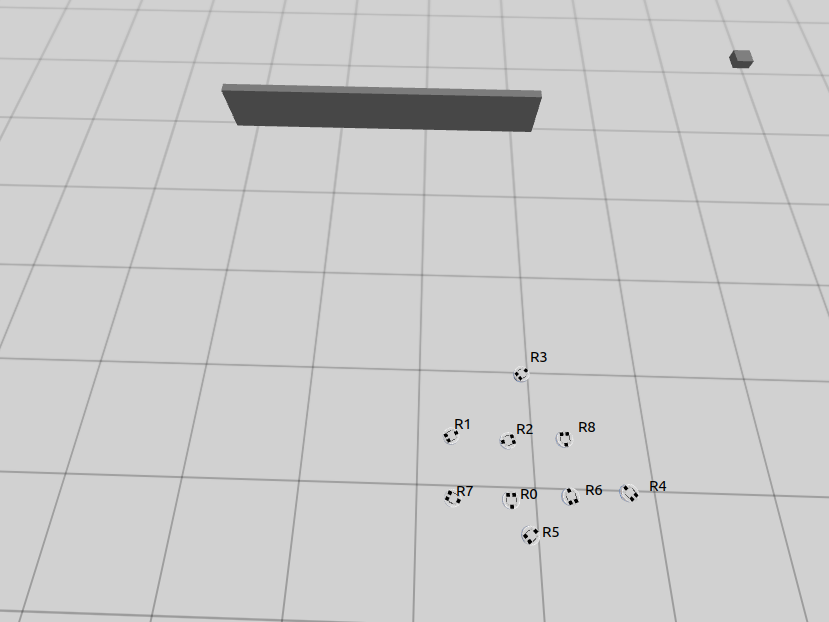}
\caption{t = 375}
\label{fig:5}
\end{figure}

\begin{figure}[htbp]
\centering
\includegraphics[width = 80mm]{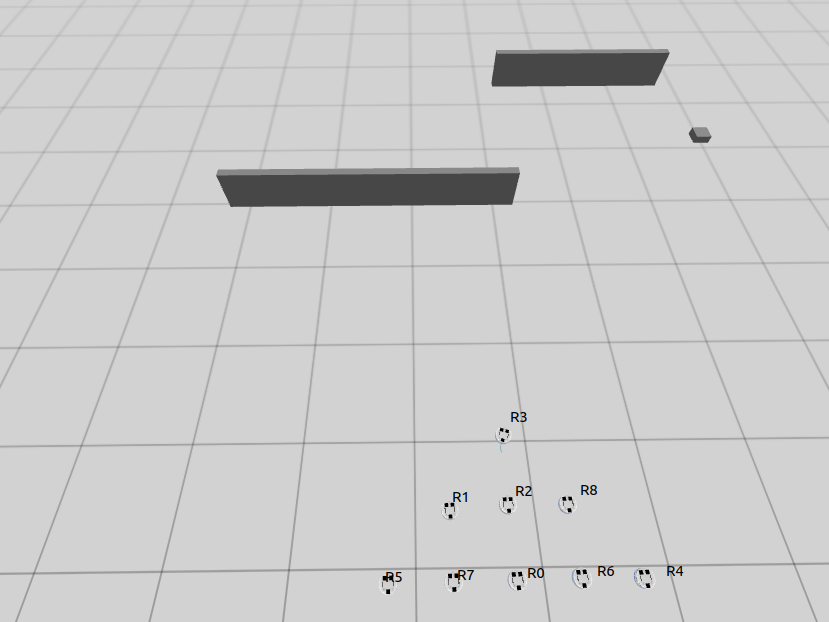}
\caption{t = 607: Motion towards destination starts}
\label{fig:6}
\centering
\includegraphics[width = 80mm]{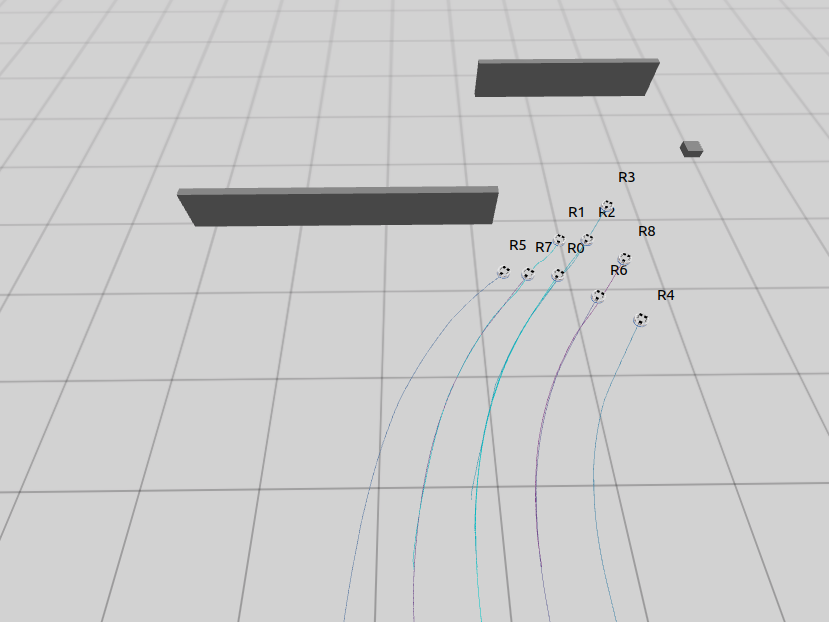}
\caption{t = 1050: The leftward perimeter avoids a large obstacle}
\label{fig:7}
\centering
\includegraphics[width = 80mm]{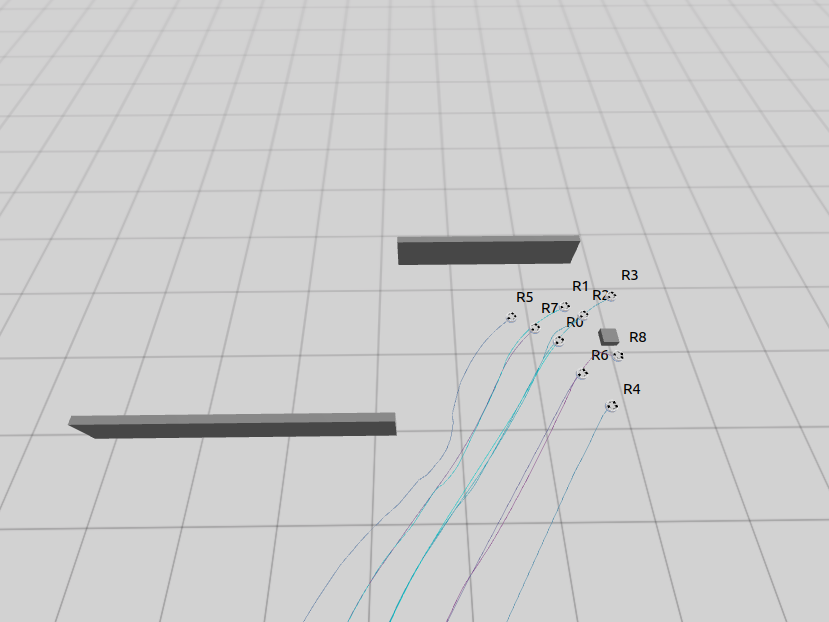}
\caption{t = 1396: The shape fragments to avoid a smaller obstacle}
\label{fig:8}
\end{figure}

\begin{figure}
\centering
\includegraphics[width = 80mm]{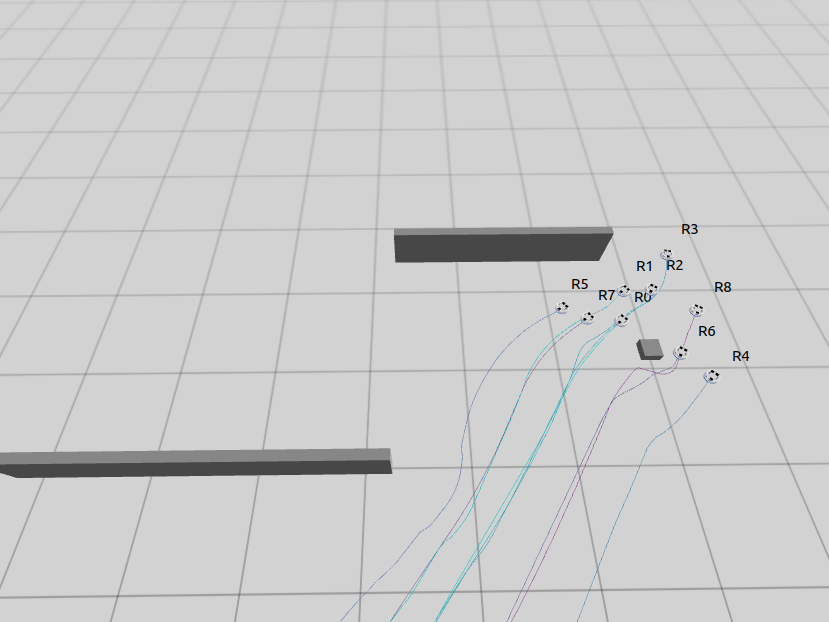}
\caption{t = 1564}
\label{fig:9}
\end{figure}

\begin{figure}[htbp]
\centerline{\includegraphics[width = 80mm]{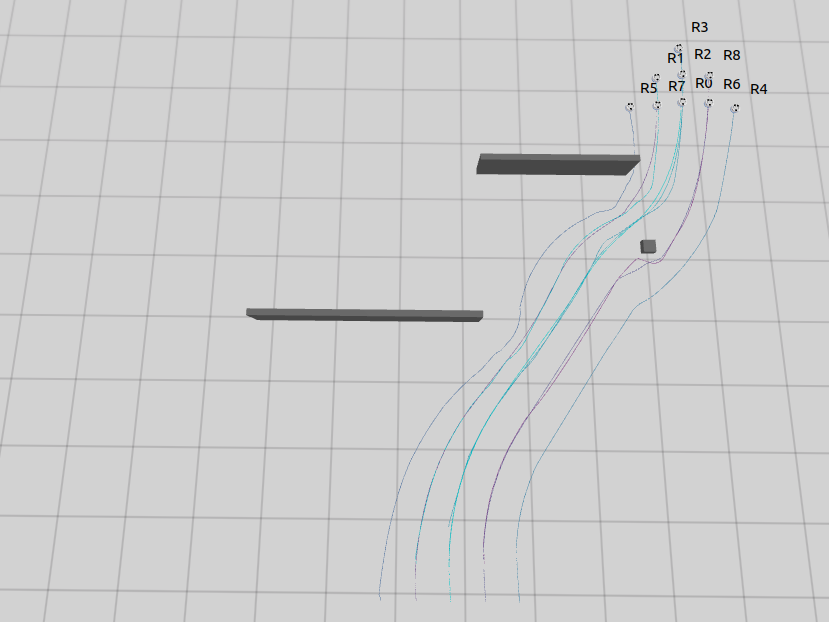}}
\caption{Trajectory of the swarm}
\label{fig:10}
\end{figure}

The experiments were conducted for various shape formations with the objective of forming the shape and moving the swarm from the origin to the destination with minimal distortion to the shape. From the residual force map plotted against the position of the centre of the formation in Fig ~\ref{fig:3}, it is inferred that the force is maximum i.e. the positional error spikes, when the swarm encounters obstacles and prioritises task of obstacle avoidance. 
The performance of the algorithm is shown in Fig ~\ref{fig:4} - ~\ref{fig:9}.
The trajectory of the bots is shown in Fig ~\ref{fig:10}
\section{Conclusion}
A decentralised approach to shape formation and a formation maintenance strategy that utilises the collective behaviour of the swarm is presented in this paper. The task switching capability and swarm-wide decision making ability is highlighted with simulation results.    
As an extension of the work, it is natural to expect that the amount of distortion would vary with the shape and size of the formation. Hence, some shapes might be more favourable to form and maintain than others. These effects can be studied for territorial exploration, search and rescue operations and object transportation where forming strategic patterns and maintaining them is important for achieving the desired results.

\section*{Acknowledgment}

We are thankful to Spider, the Research and Development club at National Institute of Technology - Tiruchirappalli for giving us the opportunity to ideate and collaborate on this project.

\vspace{12pt}
\end{document}